\documentclass{appolb}
\usepackage{graphicx}

\begin{document}
\title{Divergence of the Heavy Quasiparticle Mass at the Antiferromagnetic
Quantum Critical Point in YbRh$_2$Si$_2$%
\thanks{Presented at the Strongly Correlated Electron Systems
Conference, Krak\'ow 2002}%
}

% Authors and Affiliations

\author{P. Gegenwart, J. Custers, T. Tayama\footnote{Present address: Institute for
Solid State Physics, University of Tokyo, Kashiwa, Chiba 277-8581,
Japan}, K. Tenya\footnote{Present address: Hokkaido University,
Sapporo 060-0810, Japan}, C. Geibel, O.
Trovarelli\footnote{Present address: Infineon Technologies SC300
GmbH \& Co KG, Dresden, Germany} , and F. Steglich
\address{Max-Planck Institute for Chemical Physics of
Solids, D-01187 Dresden, Germany} \and K. Neumeier
\address{Walther Meissner Institute, D-85748 Garching, Germany}
}
\maketitle

% Abstract

\begin{abstract}
We report low temperature specific heat, $C$, magnetization, $M$,
susceptibility, $\chi$, and electrical resistivity, $\rho$,
measurements on high-quality single crystals of the heavy-fermion
system YbRh$_2$(Si$_{1-x}$Ge$_x$)$_2$ (${x=0}$ and 0.05). The
undoped compound shows weak antiferromagnetic (AF) order at
${T_N=70}$ mK which is suppressed to below 10 mK by a tiny volume
expansion in the ${x=0.05}$ system. In the latter pronounced
deviations from Landau Fermi liquid (LFL) behavior occur, e.g.
${\Delta\rho \sim T}$ over three decades in $T$. Both
thermodynamic and magnetic properties show a crossover at about
0.3 K: At 0.3 K ${\leq T \leq}$ 10 K we observe ${C/T \sim
\log(T_0/T)}$ and a "non-Curie" behavior ${\chi^{-1} \sim
T^\alpha}$ with ${\alpha<1}$ similar to what was found for the
prototypical system CeCu$_{5.9}$Au$_{0.1}$. Below 0.3 K, $\chi$
turns into a Curie-Weiss dependence ${\chi^{-1}\sim(T-\Theta)}$
indicating large unscreened Yb$^{3+}$ moments whereas in
${C(T)/T}$ a pronounced upturn occurs.\\ In the undoped compound
the AF order is suppressed continuously by critical fields
${B_{c0}\simeq 0.06}$ T and 0.7 T applied perpendicular and
parallel to the $c$-axis, respectively. For ${B>B_{c0}}$ a LFL
state with ${\Delta\rho=A(B)T^2}$ and ${C(T)/T=\gamma_0(B)}$ is
induced, that fulfills the Kadowaki-Woods scaling ${A\sim
\gamma_0^2}$. Upon reducing the magnetic field to ${B_{c0}}$ a
${1/(B-B_{c0})}$ dependence of ${A(B)}$ and ${\gamma_0^2(B)}$
indicates singular scattering at the whole Fermi surface and a
divergence of the heavy quasiparticle mass.
\end{abstract}


\section{Introduction}

The origin of non-Fermi liquid (NFL) behavior in heavy-fermion
(HF) systems has been studied intensively in the past decade but
is still unclear up to now \cite{Stewart}. In particular two
different scenarios are discussed for the quantum-critical point
(QCP), where long-range antiferromagnetic (AF) order emerges from
the HF state; one in which NFL behavior arises from Bragg
diffraction of the electrons off a critical spin-density wave
(SDW) \cite{Hertz,Millis,Moriya}, the other in which the
bound-state structure of the composite heavy fermions breaks up at
the QCP resulting in a collapse of the effective Fermi temperature
\cite{Coleman,Si}. In the SDW scenario, assuming threedimensional
(3D) spinfluctuations, singular scattering occurs only along
certain "hot lines" connected by the vector {\bf q} of the near AF
order while the remaining Fermi surface still behaves as a Fermi
liquid. Therefore, the low-temperature ($T$) specific heat
coefficient ${C(T)/T}$, that measures the heavy quasiparticle (QP)
mass, is expected to show an anomalous temperature dependence
${C(T)/T=\gamma_0-\alpha\sqrt{T}}$, but remains finite at the QCP
\cite{Moriya}. A diverging QP mass, as evident from the
${C(T)/T\sim\log(T_0/T)}$ behavior found e.g. in the prototypical
system CeCu$_{6-x}$Au$_{x}$ for ${x_c=0.1}$ \cite{Loehneysen},
would arise only if truly 2D critical spinfluctuations render the
entire Fermi surface "hot" \cite{Rosch}. On the other hand,
measurements of the inelastic neutron scattering on
CeCu$_{5.9}$Au$_{0.1}$ \cite{Schroeder} showed that the critical
component of the spin fluctuations is almost momentum independent
leading to the proposal of the locally critical scenario
\cite{Si,Schroeder}. Since $T$-dependent measurements at the QCP
alone provide no information on how the heavy quasiparticles decay
into the quantum critical state it is necessary to tune the system
away from the magnetic instability in the Landau Fermi liquid
(LFL) state and to follow the QP properties upon approaching the
QCP. In this paper we demonstrate that magnetic fields can be used
for this purpose.\\ In the following, we consider the HF metal
YbRh$_2$Si$_2$ for which pronounced NFL phenomena, i.e., a
logarithmic divergence of ${C(T)/T}$ and a quasi-linear
$T$-dependence of the electrical resistivity below 10 K, have been
observed above a low-lying AF phase transition \cite{Trovarelli
Letter}. This system is very suitable for such an investigation,
because the effect of disorder is negligible in crystals with a
residual resistivity of less than 1 ${\mu\Omega}$cm. The
application of pressure to YbRh$_2$Si$_2$ increases ${T_N}$
\cite{Trovarelli Letter} as expected, because the ionic volume of
the magnetic ${4f^{13}}$ Yb$^{3+}$-configuration is smaller than
that of the nonmagnetic ${4f^{14}}$ Yb$^{2+}$ one. Expanding the
crystal lattice by randomly substituting Ge for the smaller
isoelectric Si atoms allows one to tune
YbRh$_2$(Si$_{1-x}$Ge$_x$)$_2$ through the QCP without affecting
its electronic properties and without introducing significant
disorder to the lattice \cite{Trovarelli Physica}. In
YbRh$_2$(Si$_{1-x}$Ge$_{x}$)$_2$ with the nominal Ge concentration
${x=0.05}$ (see below) the NFL behavior extends to the lowest
accessible temperatures, in particular ${\Delta\rho(T) \sim T}$ is
observed from above 10 K down to 10 mK \cite{Trovarelli
Physica}.\\ This article is organized as follows: After giving
details concerning experimental techniques in the next section, we
address in section 3 the pronounced NFL behavior at zero magnetic
field in thermodynamic, magnetic and transport properties at the
QCP in YbRh$_2$(Si$_{0.95}$Ge$_{0.05}$)$_2$. In section 4 we
concentrate on the undoped system and prove that magnetic fields
can be used to tune the system continuously from the AF ordered
state through the QCP into a field-induced LFL state. We present
evidence for the divergence of the heavy quasiparticles  at the
QCP and end with the conclusions in section 5.

\section{Experimental details}
Single crystalline platelets of YbRh$_2$(Si$_{1-x}$Ge$_x$)$_2$
prepared with the nominal Ge compositions ${x=0}$ and 0.05 were
grown from In flux as described earlier \cite{Trovarelli
Letter,Trovarelli Physica}. A microprobe analysis (MA) of the
${x=0.05}$ sample revealed a Ge concentration ${x_{MA}\leq0.02}$,
only. The solubility of Ge atoms in In flux is much higher
compared to that of Si atoms. This causes the actual Ge
concentration being smaller than 0.05 in these single crystals. To
be consistent with previous publications \cite{Trovarelli
Physica,Custers}, the Ge-doped single crystals studied in this
paper are labeled by their nominal Ge composition ${x=0.05}$. To
demonstrate that the main effect of Ge doping is primarily the
expansion of the volume, measurements of the electrical
resistivity under hydrostatic pressure have been performed on a
${x=0.05}$ single crystal \cite{Trovarelli Physica,Mederle}. At a
pressure of 0.63 GPa the onset of AF order has been observed at
${T_N=0.185}$ K. Furthermore, the ${T_N}$ vs $p$ phase diagram of
the Ge-doped crystal matches perfectly with that found for pure
YbRh$_2$Si$_2$ \cite{Trovarelli Letter} if the pressure axis is
shifted by ${-0.2}$ GPa \cite{Mederle}. Taking the value
${B=(198\pm15)}$ GPa for the bulk modulus as determined from
recent Moessbauer measurements under hydrostatic pressure
\cite{Abd}, this pressure shift corresponds to a volume expansion
of 0.1\% for the Ge-doped sample consistent with a Ge content of
${0.02\pm0.004}$.\\ The new generation of ${x=0}$ crystals show a
residual resistivity ${\rho_0 \simeq 1 \mu\Omega}$cm, i.e., twice
as low as ${\rho_0}$ of the previous ones. Whereas for the latter
the phase transition at ${T_N}$ discovered by AC-susceptibility
measurements could not be resolved in the resistivity
\cite{Trovarelli Letter}, the new crystals show a clear kink of
${\rho(T)}$ at ${T_N}$, see below.\\ For all low-temperature
measurements, $^3$He/$^4$He dilution refrigerators were used. The
specific heat was determined with the aid of a quasi-adiabatic
heat pulse technique. The electrical resistivity, $\rho$, and the
magnetic AC-susceptibility, ${\chi}$, were measured utilizing a
Linear Research Co. (LR700) bridge at 16.67 Hz. Amplitudes of 0.1
mA and 1 Oe for the current and magnetic field, respectively, were
chosen to determine $\rho$ and ${\chi}$. Absolute values of $\chi$
have been determined from a comparison in the temperature range 2
K ${\leq T \leq}$ 6 K with the results of the dc-susceptibility in
50 mT using a Quantum Design SQUID magnetometer. Low temperature
magnetization measurements were performed utilizing a
high-resolution capacitive Faraday magnetometer as described in
\cite{Sakakibara}.

\section{Zero field NFL behavior in YbRh$_2$(Si$_{0.95}$Ge$_{0.05}$)$_2$}

\begin{figure}[!ht]
\begin{center}
\includegraphics[width=0.8\textwidth]{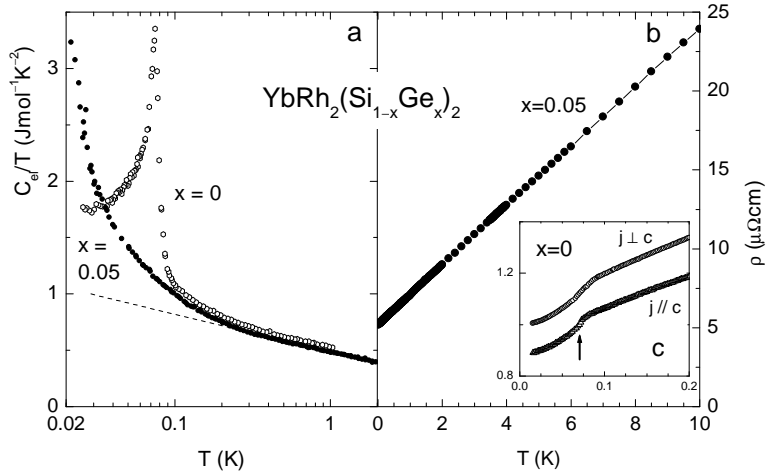}
\end{center}
\caption{Comparison of the low-temperature electronic specific
heat ${C_{el}=C-C_Q}$ as ${C_{el}/T}$ vs ${\log(T)}$ (a) and the
electrical resistivity $\rho$ (b and c) of high-quality
YbRh$_2$(Si$_{1-x}$Ge$_x$)$_2$ single crystals with Ge-contents
${x=0}$ and ${x=0.05}$. ${C_{Q}\sim T^{-2}}$ is the nuclear
quadrupole contribution calculated from recent M\"ossbauer
results~\cite{Abd}. Dotted line in (a) marks ${\log(T_0/T)}$
dependence with ${T_0\simeq~24}$ K. Resistivity data shown in (c)
were obtained for the current {\bf j} flowing along ($\Delta$) and
perpendicular ($o$) to the $c$-axis. The arrow indicates ${T_N}$
as obtained from the maximum in ${d\rho/dT}$.} \label{Fig1}
\end{figure}

\begin{figure}[!ht]
\begin{center}
\includegraphics[width=0.7\textwidth]{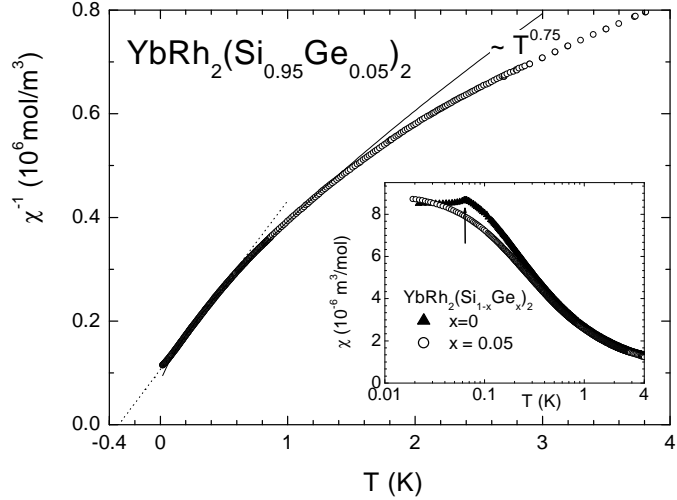}
\end{center}
\caption{Inverse of the AC-susceptibility in the easy-magnetic
plane perpendicular to the crystallographic $c$-direction, as
${\chi^{-1}}$ vs $T$ for YbRh$_2$(Si$_{0.95}$Ge$_{0.05}$)$_2$. The
dotted line indicates Curie-Weiss type dependence
${\chi(T)=C/(T-\Theta)}$ implying large fluctuating moments of 1.4
${\mu_B/Yb^{3+}}$-ion and a Weiss temperature of
${\Theta\simeq-0.3}$ K. The solid line represents ${\chi^{-1}\sim
T^\alpha}$ with ${\alpha\approx 0.75}$ (see text). The inset shows
${\chi}$ vs $T$ on a logarithmic scale for
YbRh$_2$(Si$_{1-x}$Ge$_x$)$_2$ for ${x=0}$ and 0.05. The arrow
indicates ${T_N}$ for the undoped system.} \label{Fig2}
\end{figure}

In undoped YbRh$_2$Si$_2$ we observe a second order phase
transition in specific heat (Fig. 1a) which, according to
AC-susceptibility measurements (inset of Fig. 2), marks the onset
of AF order \cite{Trovarelli Letter}. The resistivity follows a
quasi-linear $T$-dependence down to about 80 mK, below which a
sharp decrease, independent of the current direction, is observed
(inset of Fig. 1b). Thus the resistivity does not show any
signatures of the formation of a SDW for which an increase of
${\rho(T)}$ along the direction of the SDW modulation, indicating
the partial gapping of the Fermi surface, should be expected. The
absence of this behavior favors the interpretation of local-moment
type of magnetic order in YbRh$_2$Si$_2$. The resistivity in the
AF ordered state (at ${B=0}$) is best described by ${\Delta\rho
=AT^2}$ with a huge coefficient, ${A=22~\mu\Omega}$ cm/K$^2$, for
20 mK ${\leq T \leq 60}$ mK. Extrapolating ${C_{el}(T)/T}$ as
${T\rightarrow 0}$ to ${\gamma_0 = (1.7 \pm 0.2)}$ J/K$^2$mol
reveals an entropy gain at the AF phase transition of only about
${0.03R\cdot\ln2}$. This provides evidence for the weakness of the
AF order in YbRh$_2$Si$_2$. The ratio of ${A/\gamma_0^2}$ in the
ordered state is close to that expexted for a LFL \cite{KW}, i.e.,
one with very heavy quasiparticle masses.\\ The AF ordering is
suppressed to below 10 mK by doping with a small amount of Ge in
YbRh$_2$(Si$_{0.95}$Ge$_{0.05}$)$_2$. This system must be located
very close to the QCP, and pronounced NFL behavior is observed
down to lowest $T$: The electronic specific heat ${C_{el}/T}$
follows a ${\log(T_0/T)}$ (${T_0\simeq~24}$ K \cite{Trovarelli
Letter}) behavior between 0.3 and 10 K. In the same $T$-range a
linear $T$-dependence of the elecritical resistivity is observed.
Thus above 0.3 K (e.g. 0.0125 ${T_0}$) resembling $T$-dependences
as reported for CeCu$_{5.9}$Au$_{0.1}$ are observed that could
only be explained within the SDW scenario assuming truly 2D
critical spinfluctuations \cite{Rosch}. Below that temperature,
which corresponds to 75 mK in CeCu$_{5.9}$Au$_{0.1}$ (${T_0 = 6}$
K), a cross-over into a different regime takes place. Whereas the
resistivity continuous to follow a linear $T$-dependence down to
below 15 mK, a pronounced upturn is observed in ${C_{el}/T}$ whose
origin will be discussed in the conclusion. A cross-over around
0.3 K is observed in the susceptibility, too: Below 0.3 K, $\chi$
follows a Curie-Weiss type behavior (Fig. 2) implying fluctuating
moments of the order of ${1.4~\mu_B/}$ Yb$^{3+}$-ion. The
Weiss-temperature of ${\Theta \simeq -0.32}$ K suggests strong AF
correlations in this regime. Above 0.3 K, the susceptibility can
be described by ${\chi^{-1}\sim (T^\alpha-\Theta)}$ with the
exponent $\alpha$ decreasing continuously from 1 (below 0.3 K) to
0.5 at 2 K. As displayed in Fig. 2, in the $T$ interval 0.2 K
${\leq T \leq}$ 1.4 K the susceptibility roughly follows
${\chi^{-1}\sim (T^\alpha-\Theta)}$ with ${\alpha\approx 0.75}$
and ${\Theta\approx -0.05}$ K, i.e. a "non Curie-Weiss" behavior
reminiscent of CeCu$_{5.9}$Au$_{0.1}$ \cite{Schroeder}.

\section{Field tuning YbRh$_2$Si$_2$ through the QCP}

\begin{figure}[!ht]
\begin{center}
\includegraphics[width=0.65\textwidth]{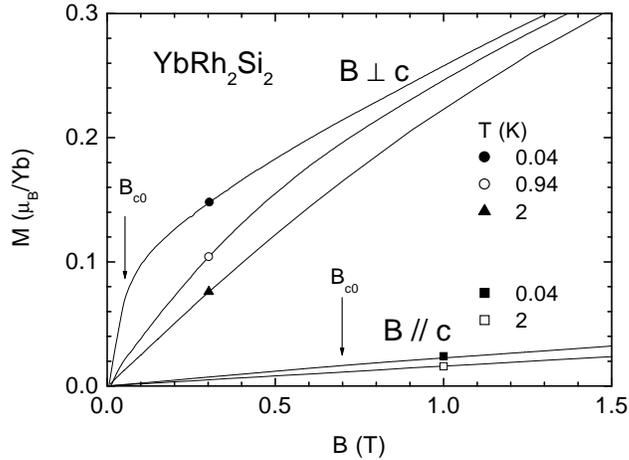}
\end{center}
\caption{Isothermal DC magnetization ${M}$ at varying temperatures
in magnetic fields applied along and perpendicular to the
$c$-axis. Arrows indicate critical fields ${B_{c0}}$.}
\label{Fig3}
\end{figure}

As discussed in the previous section, the weak AF order in
YbRh$_2$Si$_2$ can be suppressed by a tiny volume expansion upon
replacing Si by larger Ge-atoms, and pronounced deviations from
LFL behavior occur at the QCP in
YbRh$_2$(Si$_{0.95}$Ge$_{0.05}$)$_2$. To study how the heavy
quasiparticles decay into the quantum critical state it is
required to investigate their properties upon crossing the QCP at
zero temperature as a function of a control parameter that can be
varied continuously. In the following we will show that the
application of magnetic fields to YbRh$_2$Si$_2$ can be used for
this purpose. This study was motivated by the observation of
field-induced NFL behavior in the doped AF systems
CeCu$_{6-x}$Ag$_x$ \cite{Heuser98a,Heuser98b} and
YbCu$_{5-x}$Al$_x$ \cite{Seuring}. YbRh$_2$Si$_2$ is highly suited
to study the properties of a $B$-induced QCP, because the
influence of disorder in this clean stoichiometric system should
be negligible. Furthermore, the ordering temperature ${T_N = 70}$
mK is the lowest among all undoped HF systems at ambient pressure
and already a very small critical magnetic field ${B_c(0)=B_{c0}}$
is sufficient to push ${T_N}$ towards zero temperature.\\ We first
discuss the low-temperature magnetization which proves that the AF
phase transition as a function of field is a continuous one.
YbRh$_2$Si$_2$ exhibits a highly anisotropic magnetic response,
indicating that Yb$^{3+}$ moments are forming an "easy-plane"
square lattice perpendicular to the crystallographic c-direction
\cite{Trovarelli Letter}. The isothermal magnetization (Fig. 3)
shows a strongly nonlinear response for fields ${B \perp c}$. For
${T < T_N}$ a clear reduction in slope is observed above 0.06 T
which indicates the suppression of AF order resulting in a weakly
polarized state. A smooth extrapolation of ${M(B)}$ for ${B
> 0.06}$ T towards zero field reveals a value of ${\mu_s < 0.1
\mu_B}$ for the staggered magnetization in the AF state,
indicating that the size of the ordered moments is much smaller
than that of the effective moments observed in the paramagnetic
state above ${T_N}$. Thus a large fraction of the local moments
appears to remain (quantum) fluctuating within the easy plane in
the AF ordered state. Their continuous polarization for fields
exceeding ${B_{c0}}$ gives rise to a strong curvature in ${M(B)}$
for ${B \perp c}$. For fields applied along the magnetic hard
direction, ${B
\parallel c}$, the magnetization shows an almost linear behavior (Fig. 1b)
which was found to extend at least up to 58 T \cite{Custers}. At
${T < T_N}$ a very tiny increase in the ${M(B)}$ slope is observed
below about 0.7 T which, according to the resistivity measurements
discussed below, represents the critical field ${B_{c0}}$ for ${B
\parallel c}$.

\begin{figure}[!ht]
\begin{center}
\includegraphics[width=0.9\textwidth]{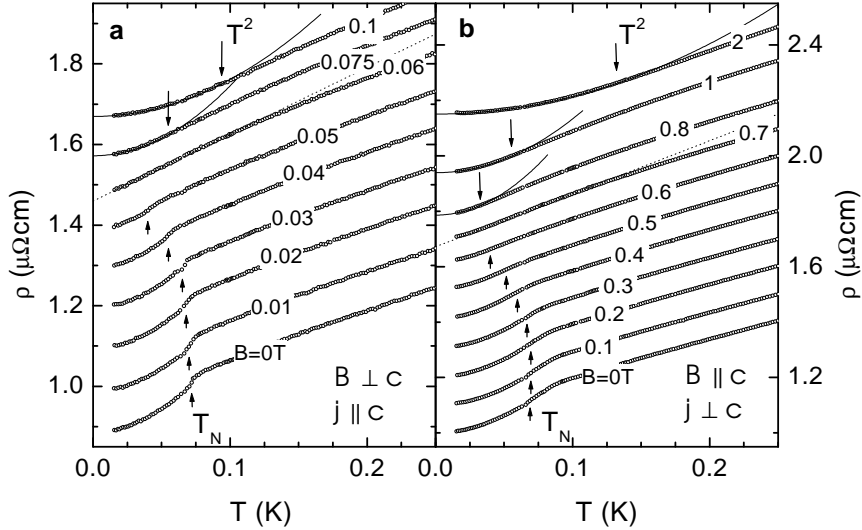}
\end{center}
\caption{Low-temperature electrical resistivity of YbRh$_2$Si$_2$
at varying magnetic fields applied perpendicular (a) and along the
$c$-axis (b). For clarity the different curves in ${B > 0}$ were
shifted subsequently by 0.1 ${\mu\Omega}$cm. Up- and downraising
arrows indicate ${T_N}$ - and upper limit of ${T^2}$ behavior,
respectively. Dotted and solid lines represent ${\Delta\rho\sim
T^\epsilon}$ with ${\epsilon=1}$ and ${\epsilon=2}$,
respectively.} \label{Fig4}
\end{figure}

\begin{figure}[!ht]
\begin{center}
\includegraphics[width=0.7\textwidth]{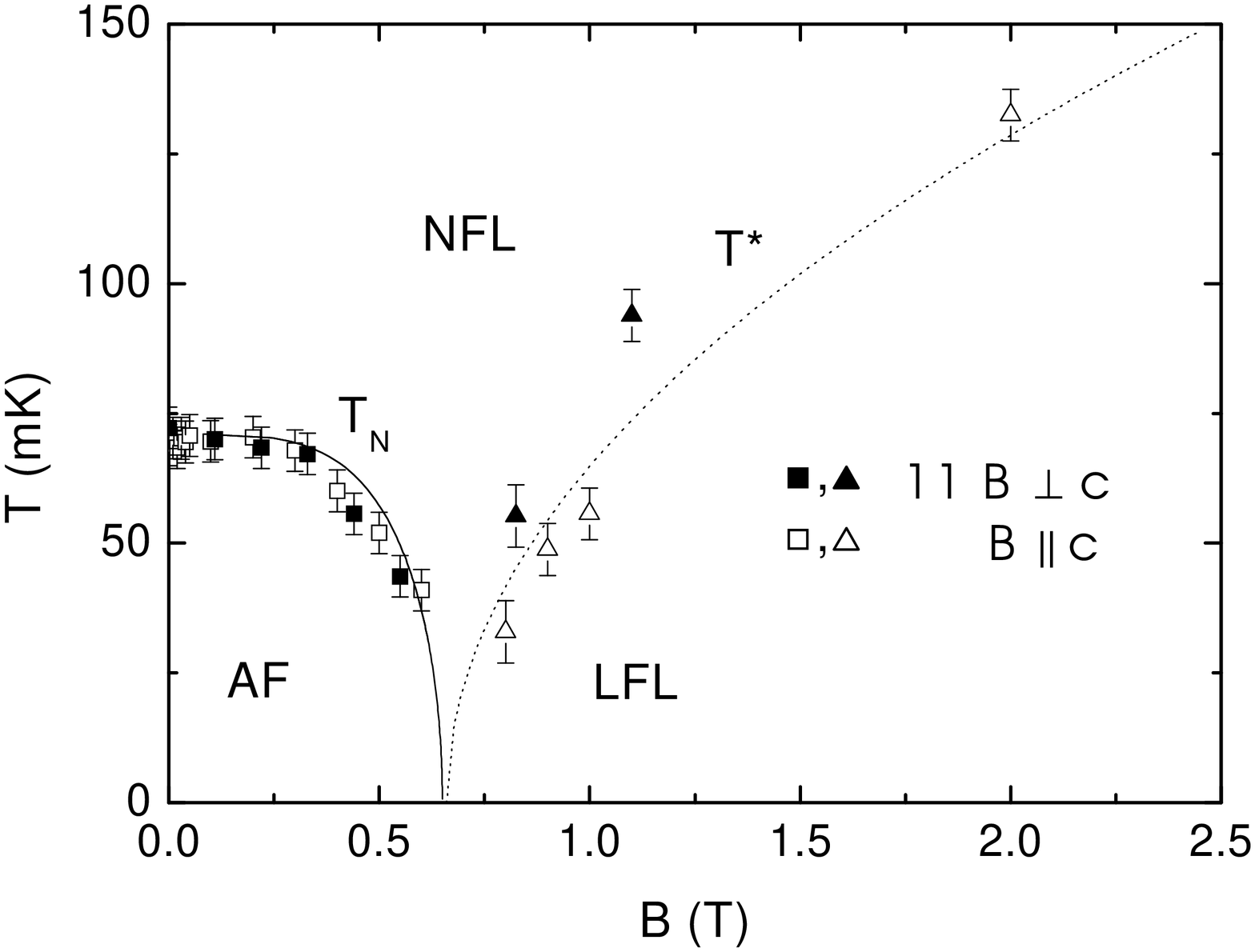}
\end{center}
\caption{${T-B}$ phase diagram for YbRh$_2$Si$_2$ with ${T_N}$ as
derived from ${d\rho/dT}$ vs $T$ and ${T^\ast}$, the  upper limit
of the ${\Delta\rho = AT^2}$ behavior, as a function of magnetic
field, applied both along and perpendicular to the $c$-axis. For
the latter ones the $B$-values have been multiplied by a factor
11. Lines separating the antiferromagnetic (AF), non-Fermi liquid
(NFL) and Landau Fermi liquid (LFL) phase are guides to the eye.}
\label{Fig5}
\end{figure}

In Fig. 4 we show the evolution of the low-temperature resistivity
upon applying magnetic fields along and perpendicular to the easy
magnetic plane. At small magnetic fields the N\'{e}el temperature,
determined from the maximum value of ${d\rho/dT}$, shifts to lower
temperatures and vanishes at a critical magnetic field ${B_{c0}}$
of 0.06 T in the easy magnetic plane and of 0.66 T along the
magnetic hard direction, i.e. the crystallographic $c$-axis. At
${B = B_{c0}}$, the resistivity follows a linear $T$-dependence
down to the lowest accessible temperature of about 20 mK. This
observation provides striking evidence for field-induced NFL
behavior at the critical magnetic fields applied along both
crystallographic directions \cite{Gegenwart}. At ${B>B_{c0}}$, we
find ${\Delta\rho = A(B)\cdot T^2}$ for ${T \leq T^\ast(B)}$, with
${T^\ast (B)}$ increasing and ${A(B)}$ decreasing upon raising the
applied magnetic field. The evolution of ${T_N}$ and ${T^\ast}$ as
a function of $B$ is shown in Fig. 5. The extremely low value of
the critical field applied along the easy plane highlights the
near degeneracy of two different heavy LFL states, one being
weakly AF ordered (${B<B_{c0}}$) and the other one being weakly
polarized (${B>B_{c0}}$).

\begin{figure}[!ht]
\begin{center}
\includegraphics[width=0.75\textwidth]{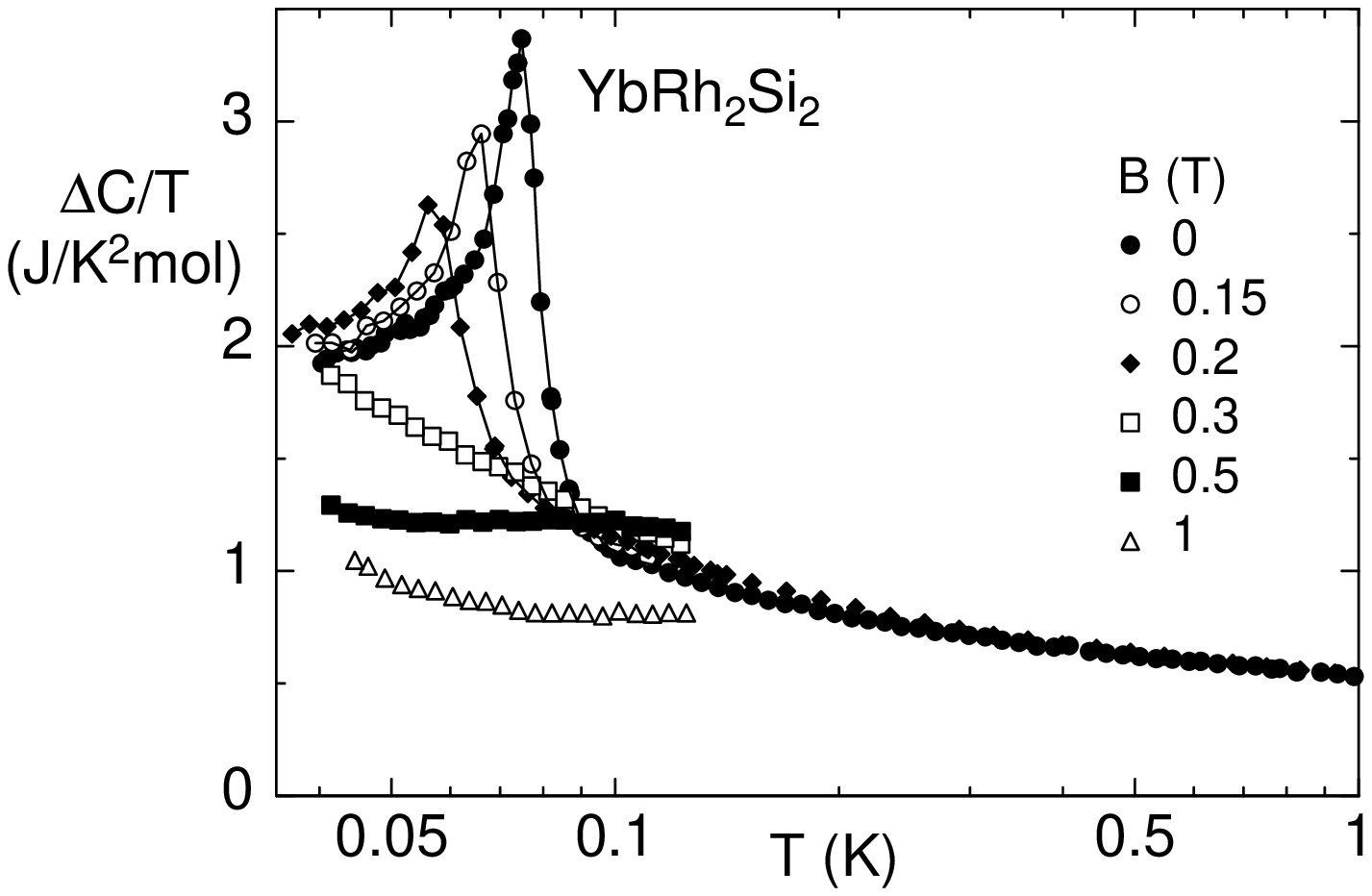}
\end{center}
\caption{Specific heat as ${\Delta C/T=(C-C_Q)/T}$ vs $T$ (on a
logarithmic scale) for YbRh$_2$Si$_2$ at varying fields applied
parallel to the $c$-axis. ${C_Q \sim T^{-2}}$ is the nuclear
quadrupole contribution calculated from recent Moessbauer
results\cite{Abd}.} \label{Fig6}
\end{figure}

Next we address specific-heat measurements (Fig. 6) which have
been performed at magnetic fields along the $c$-axis. Due to the
strong magnetic anisotropy, the sample plate used for the
specific-heat measurement could not be aligned perfectly along the
hard magnetic direction. Therefore, a critical field ${B_{c0}}$ of
only about 0.3 T was sufficient to suppress AF order completely in
these experiments. As shown in Fig. 6, at ${B = B_{c0}}$ the
specific-heat coefficient ${\Delta C(T)/T}$ increases down to the
lowest $T$, indicative of a field-induced NFL ground state. Within
40 mK ${\leq T \leq 120}$ mK it follows a steep increase similar
to the upturn observed for the ${x=0.05}$ single crystal below 0.3
K (c.f. Fig. 1a). While this anomalous contribution is strongly
reduced upon increasing $B$, at magnetic fields ${B \geq}$ 1 T,
the nuclear contribution becomes visible at the lowest
temperatures, above which a constant ${\gamma_0(B)}$ value is
observed within a limited temperature window (Fig. 6).
${\gamma_0(B)}$ decreases in magnitude upon increasing the
field.\\A very similar behavior has been found in the field
dependence of the low temperature AC susceptibility, too
\cite{Trovarelli Letter}. For magnetic fields exceeding ${B_{c0}}$
constant values ${\chi_0(B)}$ have been observed which decrease in
size upon increasing $B$.

\begin{figure}[!ht]
\begin{center}
\includegraphics[width=0.9\textwidth]{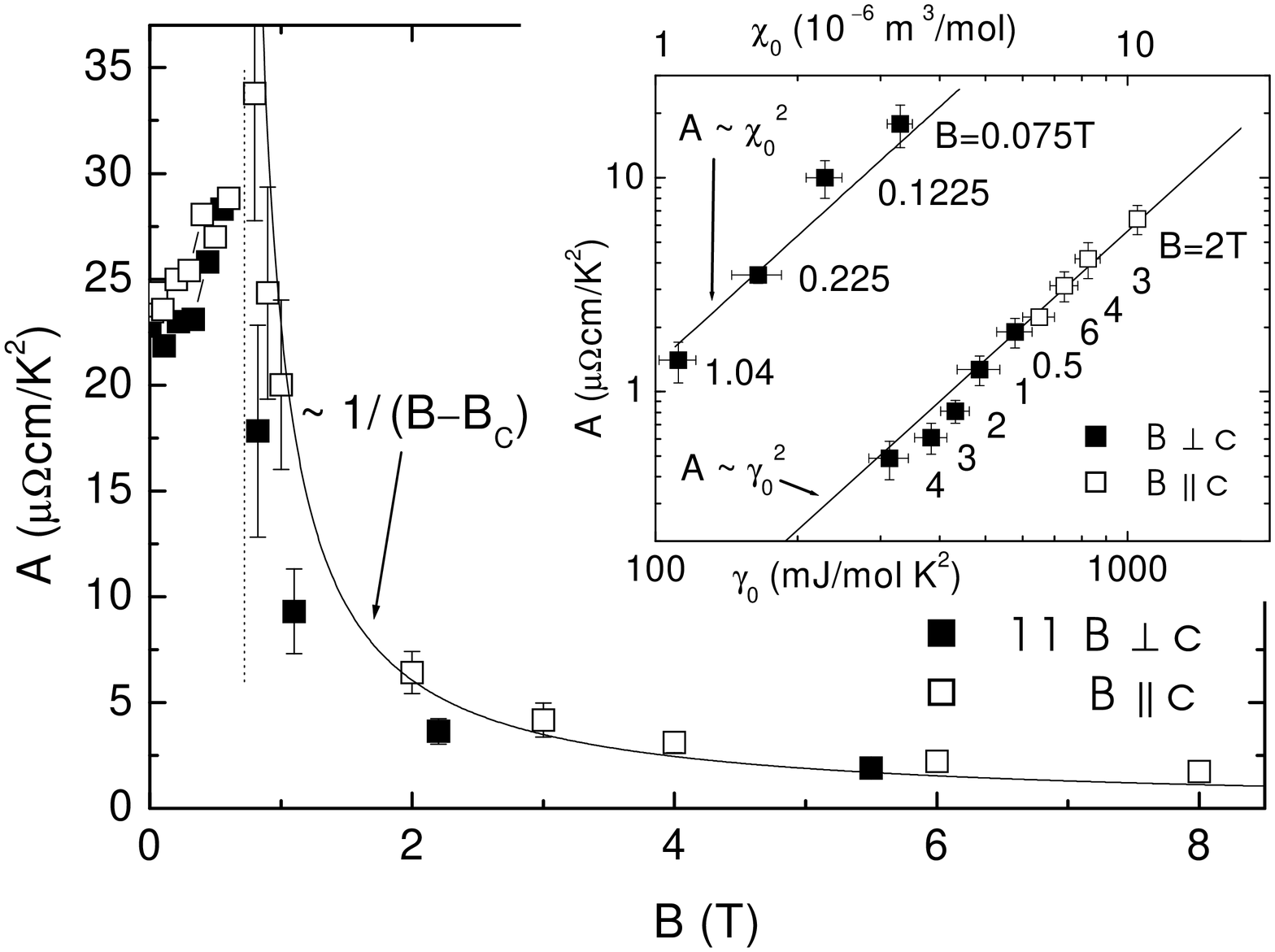}
\end{center}
\caption{Coefficient ${A = \Delta\rho/T^2}$ vs field $B$. Data for
$B$ perpendicular to the $c$-direction have been multiplied by 11.
Dashed line marks ${B_{c0}}$, solid line represents
${(B-B_{c0})^{-1}}$. Inset shows double-$\log$ plot of $A$ vs
${\gamma_0}$ and $A$ vs ${\chi_0}$ for different magnetic fields.
Solid lines represent ${A/\gamma_0^2 = 5.8\cdot10^{-6}
\mu\Omega}$cm(Kmol/mJ)$^2$ and ${A/\chi_0^2 = 1.25\cdot10^{12}
\mu\Omega}$cmK$^{-2}$/(m$^3$/mol)$^2$.} \label{Fig7}
\end{figure}

In Fig. 7 we present our analysis of the magnetic-field dependence
of the coefficients $A$, ${\gamma_0}$ and ${\chi_0}$ observed for
${T \rightarrow 0}$ in the resistivity, ${\Delta\rho = A(B)T^2}$,
specific heat, ${\Delta C/T=\gamma_0(B)}$ \cite{Trovarelli
Letter}, and magnetic AC-susceptibility, ${\chi=\chi_0(B)}$
\cite{Trovarelli Letter} when approaching the QCP upon reducing
$B$ towards ${B_{c0}}$. As shown in the inset of Fig. 7, we
observe ${A\sim \gamma_0^2, A \sim \chi_0^2}$ and thus also
${\gamma_0 \sim \chi_0}$, independent of the field orientation and
for all $B$ values exceeding ${B_{c0}}$. Like in the AF ordered
state (at ${B=0}$) we find that the ${A/\gamma_0^2}$ ratio roughly
equals that observed for many HF systems \cite{KW}. A very large
Sommerfeld-Wilson ratio ${R =
(\chi_0/\gamma_0)(\pi^2k_B^2/\mu_0\mu_{eff}^2)}$ of about 14
${(\mu_{eff} = 1.4 \mu_B)}$ indicates a strongly enhanced
susceptibility in the field-aligned state of YbRh$_2$Si$_2$
pointing to the importance of low-lying ferromagnetic (${\bf
q=0}$) fluctuations, as also inferred from recent $^{29}$Si-NMR
measurements \cite{Ishida}.\\Since YbRh$_2$Si$_2$ behaves as a
true LFL for ${B > B_{c0}}$ and ${T < T^\ast(B)}$, the observed
temperature dependences should hold down to ${T = 0}$. The field
dependence ${A(B)}$ shown in Fig. 7 measures the QP-QP scattering
cross section when, by field tuning, crossing the QCP at zero
temperature. A ${1/(B-B_{c0})}$ divergence is observed indicating
that the whole Fermi surface undergoes singular scattering at the
$B$-tuned QCP. Most importantly, the relation ${A \sim
\gamma_0^2}$ observed at elevated fields, suggests that also the
QP mass diverges, i.e., as ${1/(B-B_{c0})^{1/2}}$, upon
approaching $B_{c0}$.

\section{Conclusion}

The HF metal YbRh$_2$Si$_2$ shows a weakly antiferromagnetically
polarized ground state which is suppressed to ${T_N\rightarrow~0}$
either by a small volume expansion in
YbRh$_2$(Si$_{0.95}$Ge$_{0.05}$)$_2$ or by the application of a
critical magnetic field ${B_{c0}}$. At the QCP, pronounced
deviations from LFL behavior are observed below 10 K in the
specific heat and the electrical resistivity, in particular
${C/T\sim\log(T_0/T)}$ (above 0.3 K) and ${\Delta\rho\sim~T}$
(down to the lowest $T$). According to the SDW scenario, such
$T$-dependences would only arise assuming the scattering of truly
2D critical spinfluctuations.\\ In order to study the decay of the
heavy quasiparticles in the quantum-critical state, we have used
magnetic fields to suppress the AF order in YbRh$_2$Si$_2$. By
field tuning the system through the QCP at a critical field
${B_{c0}}$ one reaches a field-aligned state which at low
temperatures can be described by the LFL model. The
${1/(B-B_{c0})}$ divergence of $A(B)$ and ${\gamma_0^2(B)}$
indicates a divergence of the QP mass and QP-QP scattering cross
section upon approaching the QCP. In the SDW model the parameter
$\delta$, given by the square of the inverse magnetic correlation
length in the 2D spin fluid, measures the distance from the QCP,
i.e., ${\delta\sim(B-B_{c0})}$. Assuming that the spin fluid
renders the entire Fermi surface "hot", the coefficient $A$
diverges as ${A\sim1/\delta}$, whereas for the specific heat
coefficient ${\gamma_0}$ a much weaker divergence ${\gamma_0 \sim\
log(1/\delta)}$ is expected \cite{Kotliar}. Thus, this model would
predict the ratio ${A/\gamma_0^2}$ to diverge upon decreasing $B$
instead of being constant. The constancy of the Kadowaki-Woods
ratio when approaching the QCP rather favors the locally critical
scenario \cite{Piers}.\\ A stronger than logarithmic divergence of
the QP mass is also evident from our ${B=0}$ measurements of
${C_{el}(T)/T}$ for YbRh$_2$(Si$_{0.95}$Ge$_{0.05}$)$_2$ below 0.3
K(Fig. 1a). The upturn cannot be explained assuming a nuclear
contribution to the low-$T$ specific heat. Furthermore a very
similar behavior is observed for the volume thermal expansion
coefficient $\beta$ too: As found for ${C_{el}/T}$, above 0.3 K
${\beta/T}$ follows a logarithmic $T$ dependence \cite{Oeschler}.
Below that temperature an even stronger divergence has been
observed, providing additional evidence for the electronic origin
of the upturn in the specific-heat coefficient. The Curie-Weiss
behavior, observed in ${\chi(T)}$ in the same $T$-range, where the
upturn in ${C_{el}(T)/T}$ occurs, hints to large unscreened
fluctuating Yb$^{3+}$ moments persisting down to the lowest $T$ at
the QCP. This strongly suggests a local nature of the critical
fluctuations.\\ To conclude, the QP mass diverges faster than
logarithmic upon approaching the QCP in YbRh$_2$Si$_2$ in
contradiction to the SDW picture and suggests a locally critical
scenario for this system.

We gratefully acknowledge stimulating discussions with Catherine
Pepin, Piers Coleman, Qimiao Si and Heribert Wilhelm. This work
was supported by the FERLIN program of the ESF.

\end{document}